\documentclass[a4paper]{jpconf}
\usepackage{hyperref}
\usepackage{graphicx}
\pdfoutput=1
\begin{document}
\title{Latest results on differential cross sections as a function of kinematic variables at the LHC and Tevatron}

\author{Carmen Diez Pardos \\ for the ATLAS, CDF, CMS and D0 collaborations}

\address{DESY, Notkestr. 85 22607, Hamburg, Germany}

\ead{carmen.diez@desy.de}

\begin{abstract}
An overview of recent measurements of differential top quark pair ($t\bar{t}$) production cross sections is presented, including results in boosted topologies and results both at parton and particle level. The first measurements of differential single top quark t-channel production cross sections in proton-proton collisions are as well presented. The results are obtained using data collected with the ATLAS and CMS experiments in proton-proton collisions at centre-of-mass energies of 7 TeV and 8 TeV. References to the latest results by the Tevatron experiments are also given. The data are compared with several predictions from perturbative QCD calculations up to approximate next-to-next-to-leading-order precision (approx. NNLO). 

\end{abstract}

\section{Introduction}
Measurements of differential cross sections as a function of top quark related kinematic quantities are tests of perturbative QCD (pQCD) and probe a variety of different properties, for instance the $p_{t\bar{t}}$ distribution is sensitive to higher order effects like initial/final state radiation of quarks or gluons. Moreover, possible deviations in the shapes of all distributions can provide a hint of physics beyond the SM.

Top quarks are mostly produced in pairs via the strong interaction in hadron colliders. At LHC energies, the dominant mechanism is gluon-gluon fusion, corresponding to $\sim$80\% of the generation process. Top quarks can also be produced singly, via the electroweak interaction. The three modes of single top quark production are t-channel, tW-associated production and s-channel. The top quark decays almost exclusively into a W boson and a b quark and it is the decay of the W bosons what defines the final state. Therefore, $t\bar{t}$ signatures can be classified according to the combinatorics of the W boson decay. The $t\bar{t}$ measurements presented here are performed in the dilepton channel, where the $t\bar{t}$ final states include events with two leptons, two neutrinos and two b jets, and/or in the lepton+jets channels, with one lepton, one neutrino and four jets, out of which two arise from b quarks. The t-channel single top quark analyses consider only the leptonic decay of the top quark.
 
In the following the latest analyses of top quark production performed at CMS~\cite{CMSColl} and ATLAS~\cite{AtlasColl} are summarised, focusing on the measurements obtained with data collected at 8~TeV when available. Details of the latest results by the CDF and D0 collaborations and results with data collected at 7~TeV by CMS, not described here, can be found in~\cite{CDF},~\cite{D0} and~\cite{CMS7TeV}, respectively.

\section{Differential $t\bar{t}$ production cross section}
\begin{figure}[htbp!]
  \begin{center}
\includegraphics[width=0.33\textwidth]{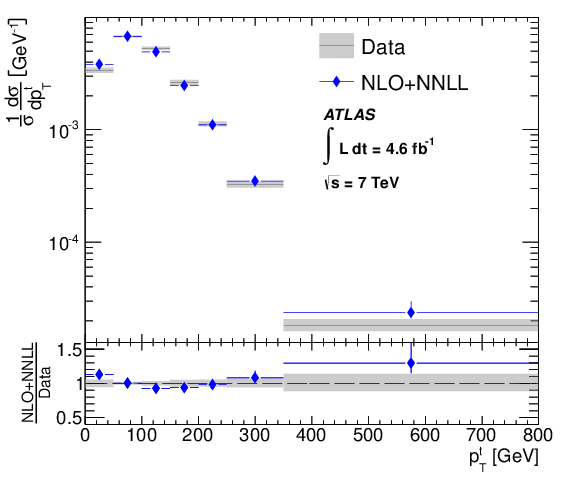}%
\includegraphics[width=0.33\textwidth]{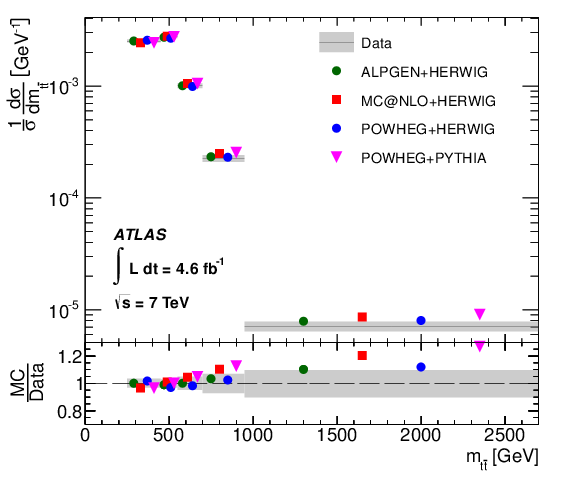}%
      \includegraphics[width=0.33\textwidth]{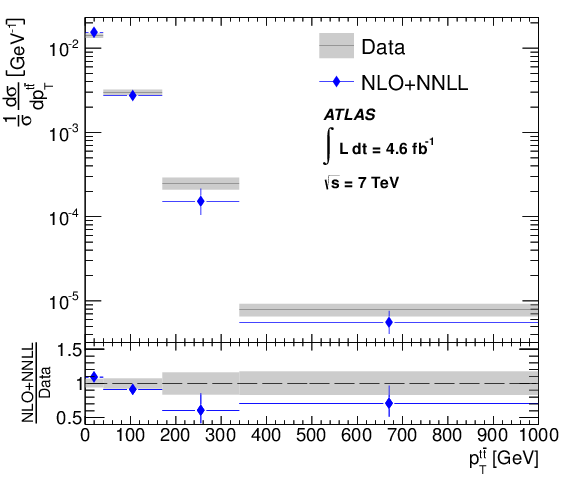}\\
\includegraphics[width=0.33\textwidth]{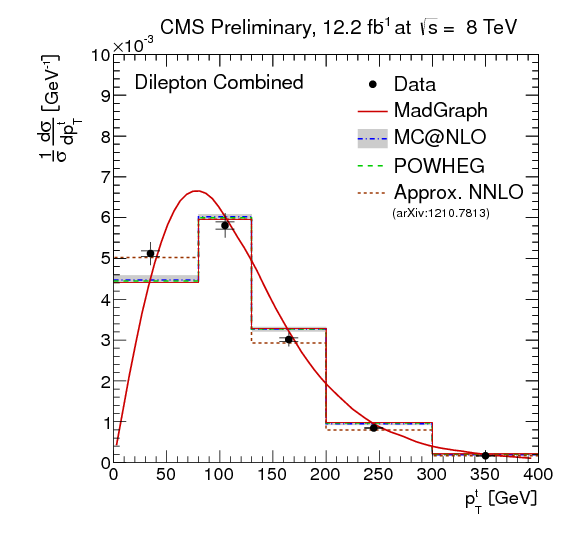}%
\includegraphics[width=0.33\textwidth]{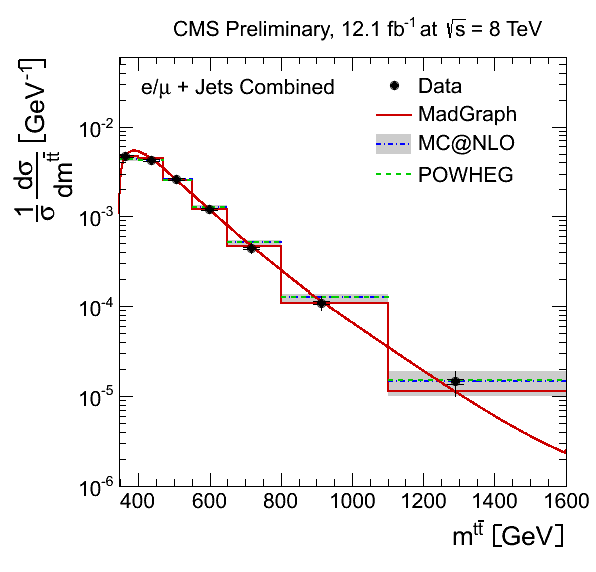}%
\includegraphics[width=0.33\textwidth]{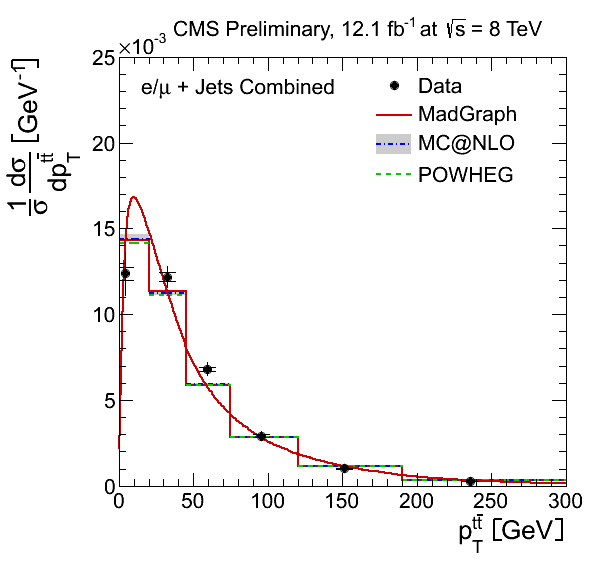}
  \end{center}
  \caption{Normalised differential $t\bar{t}$ production cross section as a function of p$_T^t$ (left), $m_{t\bar{t}}$ (middle) and p$_T^{t\bar{t}}$ (right), measured by ATLAS (top)~\cite{ATLASljets} and CMS (bottom)~\cite{CMSljets,CMSdilepton}. The band (top) and outer bars (bottom) represent the total uncertainty.}
  \label{fig:topxsec}
\end{figure}

The measurements presented in this report are performed in the lepton+jets and/or the dilepton decay channels. The top quark kinematic properties are obtained through kinematic fitting and reconstruction algorithms. The measurements are corrected for detector and hadronisation effects back to particle or parton level, using a regularised unfolding procedure. Theory predictions and calculations are compared to the differential cross sections, absolute or normalised to the in-situ measured absolute cross section, $\sigma$, extrapolated to the full phase space or in the visible phase space, in order to minimise the dependence on theory input. The measurements have been performed as a function of several kinematic distributions of top quark, $t\bar{t}$, (b)-jets, leptons, lepton pairs, etc. Here, few examples of results for variables related to the top quark and $t\bar{t}$ are discussed. 
The normalised cross sections as a function of the top quark transverse momentum, p$_T^t$, the invariant mass, m$_{t\bar{t}}$, and p$_T$ of the $t\bar{t}$, p$_T^{t\bar{t}}$, are presented in Fig.~\ref{fig:topxsec}. The results are obtained in the lepton+jets channels~\cite{CMSljets,ATLASljets} and dilepton channels~\cite{CMSdilepton}. The measurement uncertainties range typically between 2-3\% up to 20\% and are generally dominated by systematic effects. In general, the predictions and the NLO+NNLL and approx. NNLO calculations for m$_tt$ and p$_T^t$, respectively, are softer than the data for both experiments, and the prediction by POWHEG+HERWIG provides the best description of data. The NLO+NNLL calculation does not describe p$_T^t$ well.
Results for the directly measurable quantities, such as the kinematic properties of leptons and b jets, obtained in a visible phase defined by the kinematic and geometrical acceptance of all selected final-state objects can be found in~\cite{CMS7TeV,CMSljets,CMSdilepton}.

\begin{figure}[htbp!]
  \begin{center}
      \includegraphics[width=0.45\textwidth]{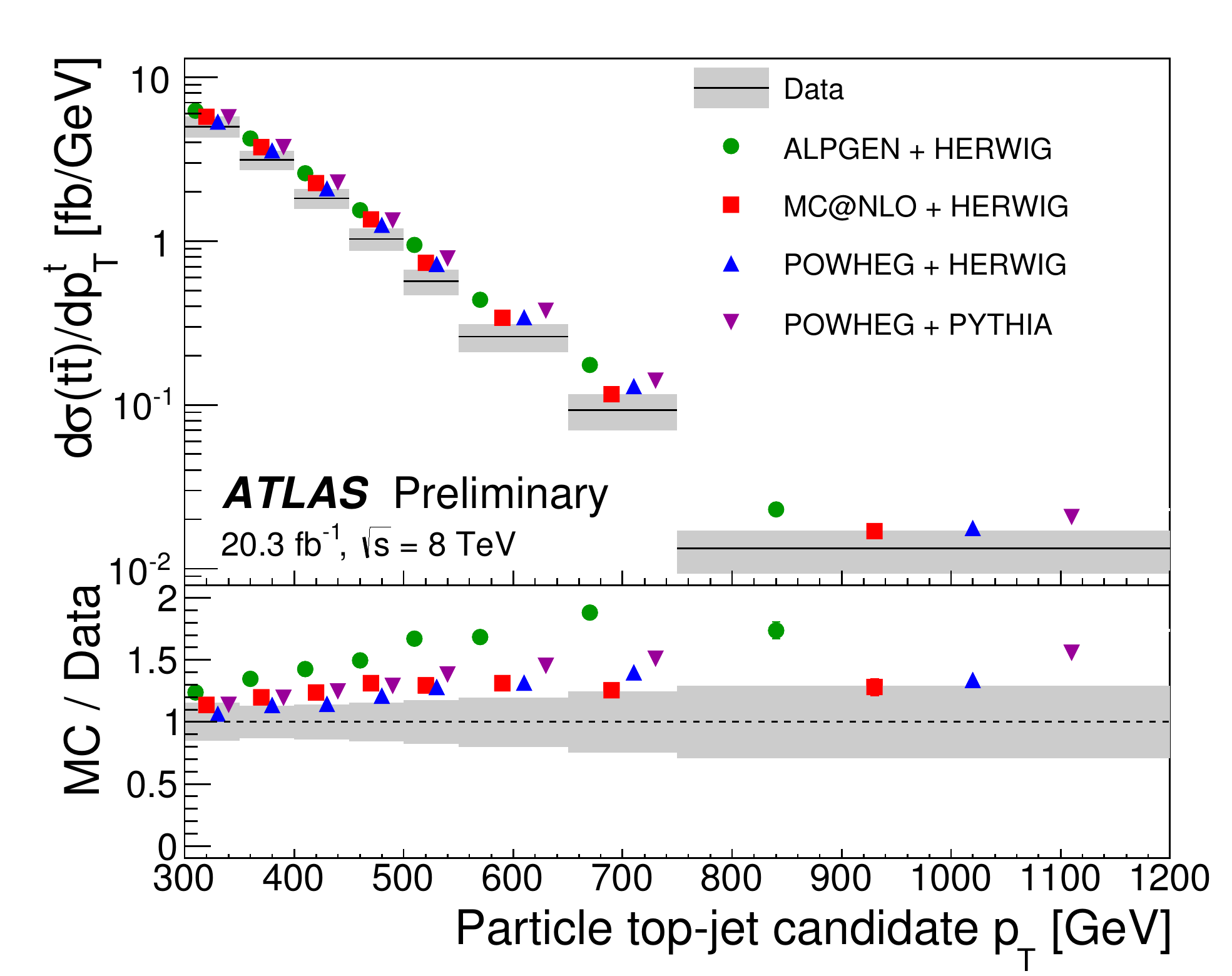}%
      \includegraphics[width=0.45\textwidth]{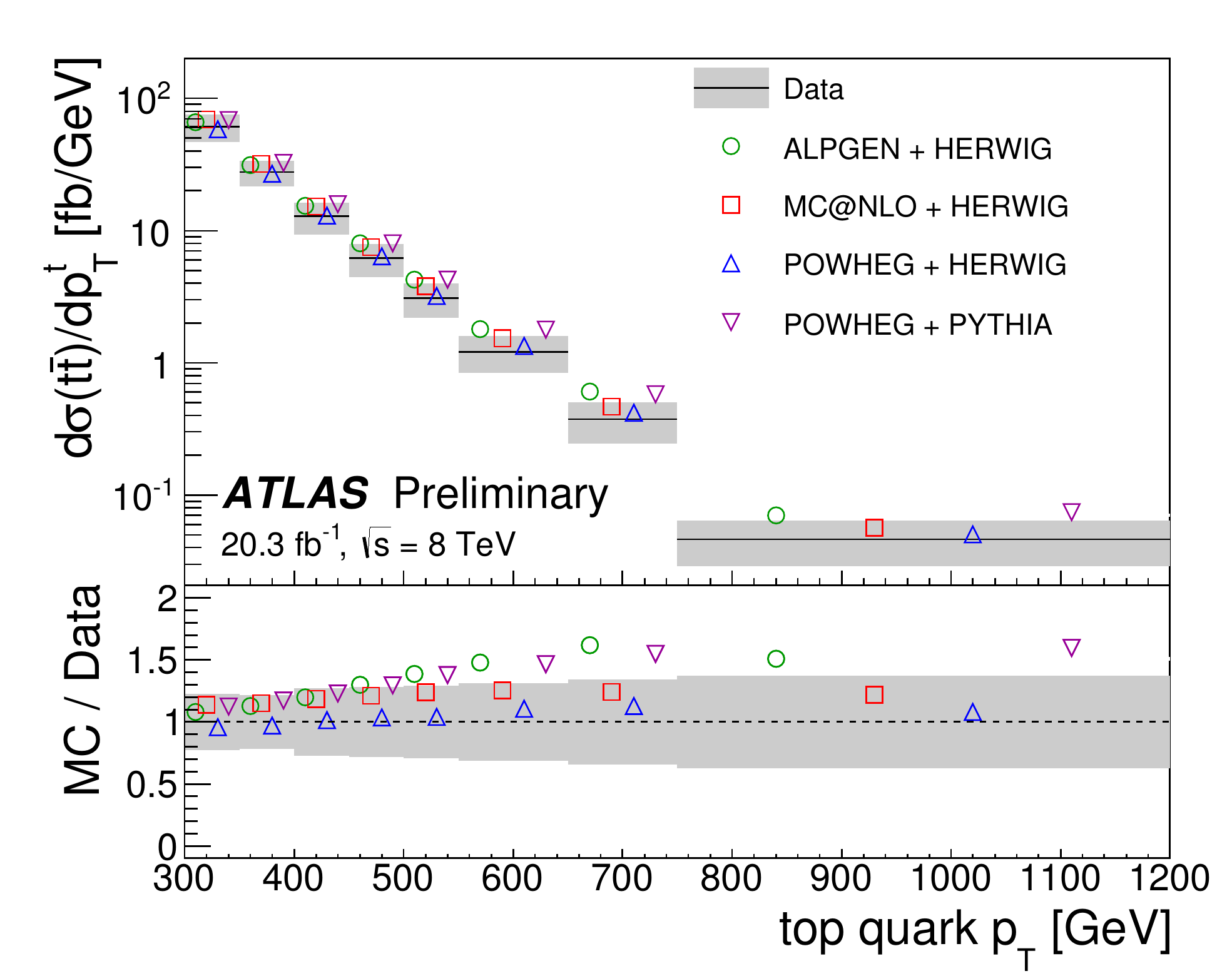}     
      \caption{Fiducial particle-level differential cross section as a function of the hadronic top-jet candidate p$_T$ (left) and parton-level result (right) as a function of the top quark p$_T$ decaying hadronically~\cite{ATLASboosted}. The shaded area corresponds to the total uncertainty.}
      \label{fig:boosted}
  \end{center}
\end{figure}

A dedicated analysis has been performed by ATLAS in the lepton+jets channel to measure the absolute differential cross section for boosted $t\bar{t}$ production. The results are obtained as a function of the hadronically decaying top quark transverse momentum~\cite{ATLASboosted}, with p$_T$ larger than 300~GeV. Jet substructure techniques are employed to identify top quarks, which are reconstructed with an anti-kt jet with radius parameter R = 1.0. The results are presented at particle level in the visible phase space, close to the event selection, and extrapolated at parton level to the full phase space, up to the TeV scale for top quarks with transverse momentum above 300~GeV, see Fig.~\ref{fig:boosted}. The total uncertainties range between 15\% (20\%) and 20\% (40\%) at particle level (parton level). The predictions are found to generally overestimate the measured cross sections, the discrepancy increases with p$_T$.

\begin{figure}[htbp!]
  \begin{center}
      \includegraphics[width=0.35\textwidth]{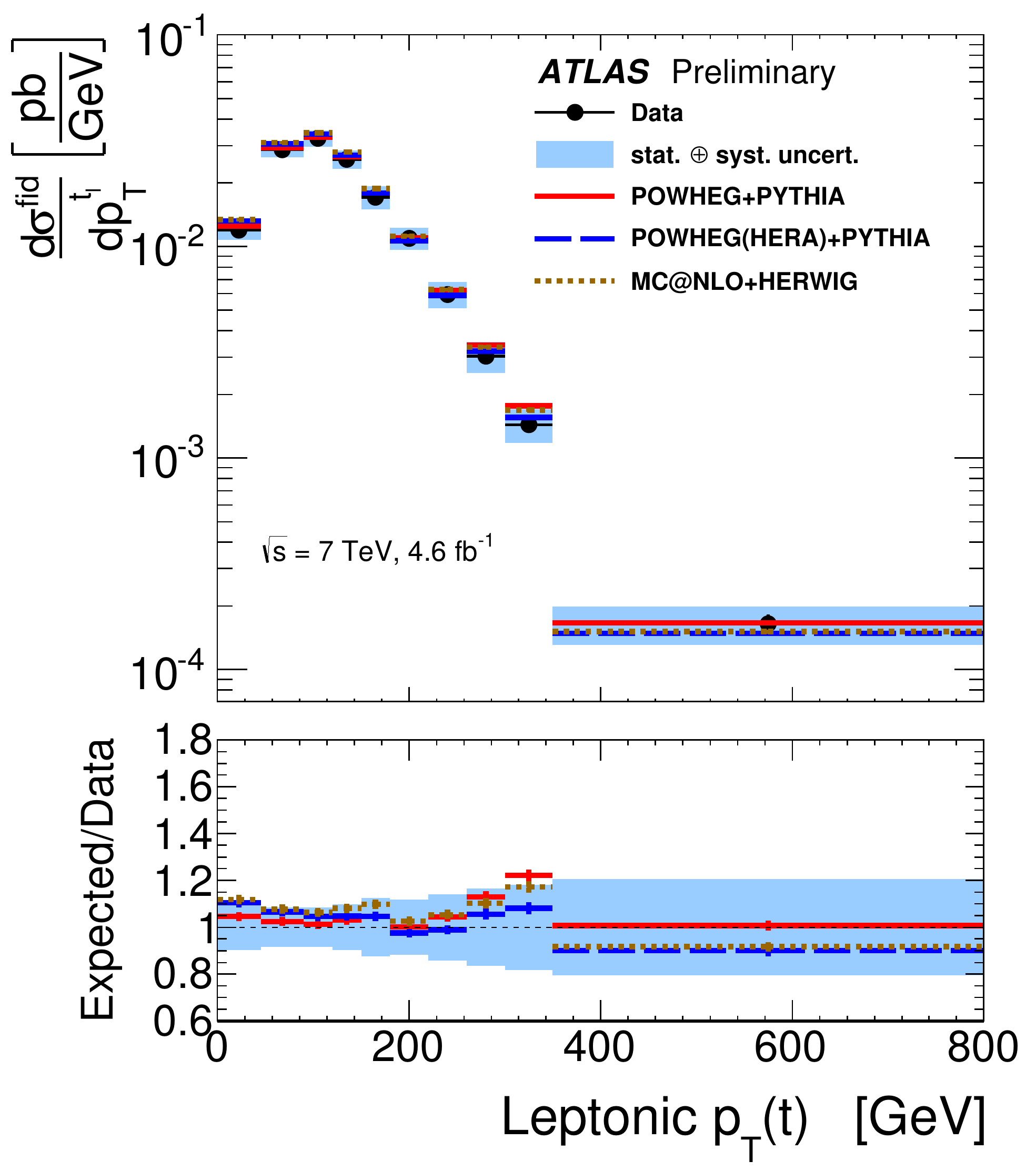}  
      \includegraphics[width=0.35\textwidth]{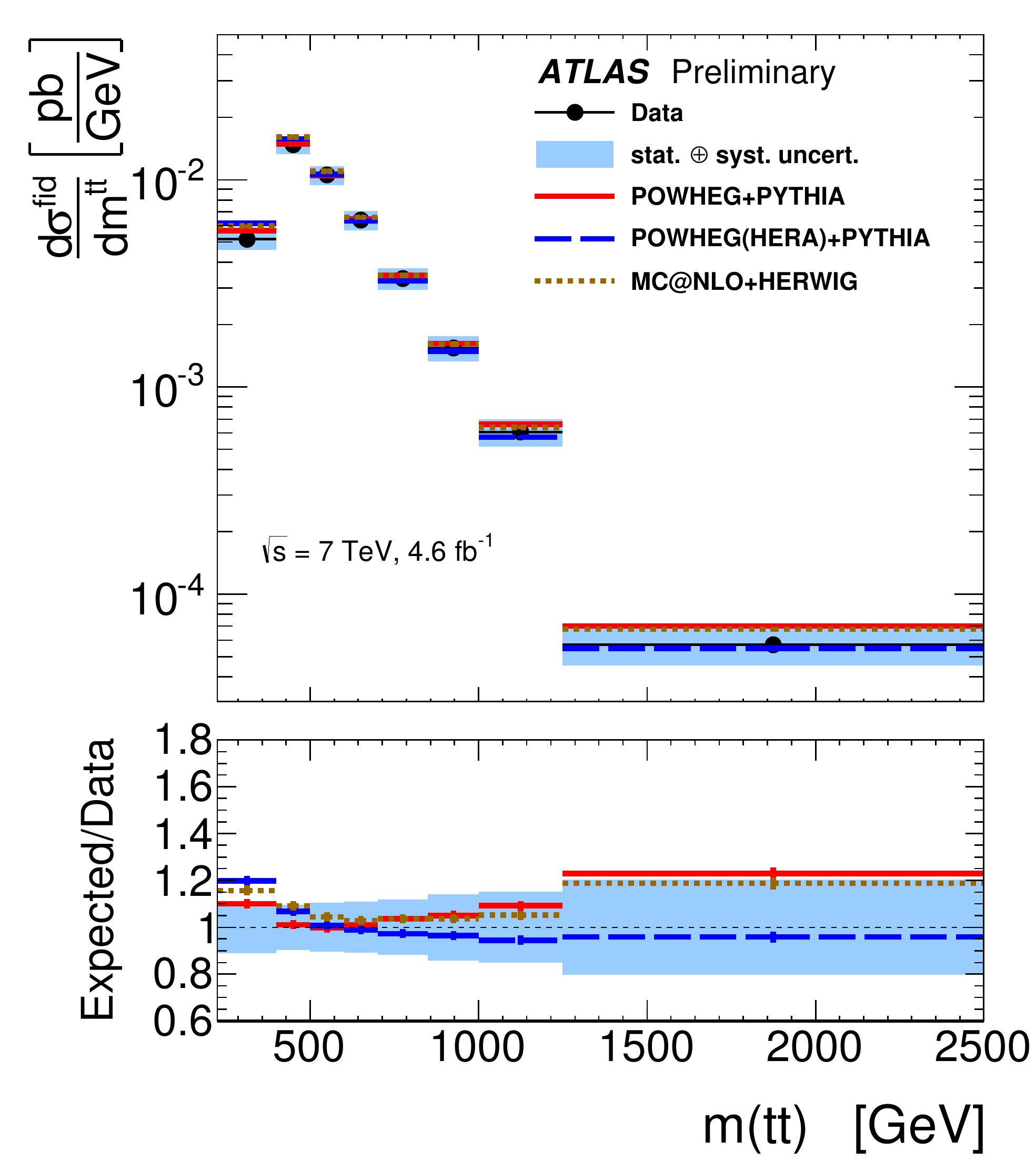}
      \caption{Differential $t\bar{t}$ cross section as a function of the leptonic pseudo-top-quark p$_T$ (left) and $m_{tt}$ (right)~\cite{ATLASpseudo}. The data points are shown with a blue band which represents the total uncertainty.}
      \label{fig:pseudo}
  \end{center}
\end{figure}

The differential $t\bar{t}$ production cross section has been recently measured in terms of an observable referred to as the pseudo-top-quark, where the differential variables are constructed from objects that are directly related to detector level observables (jets, leptons and missing transverse momentum)~\cite{ATLASpseudo}. This observable is defined both in terms of reconstructed objects and stable particles. The measurements are presented after corrections for detector effects, within a kinematic range that closely matches the detector acceptance to minimise model-
dependent extrapolations of the data. The absolute production cross section as a function of the transverse momentum of the leptonic pseudo-top-quark, as well as the invariant mass of the pseudo-top-quark pair system are shown in Fig.~\ref{fig:pseudo}, compared to the predictions from several NLO MC generators.

\section{Differential single top production cross section}
\begin{figure}[htbp!]
  \begin{center}
       \includegraphics[width=0.33\textwidth]{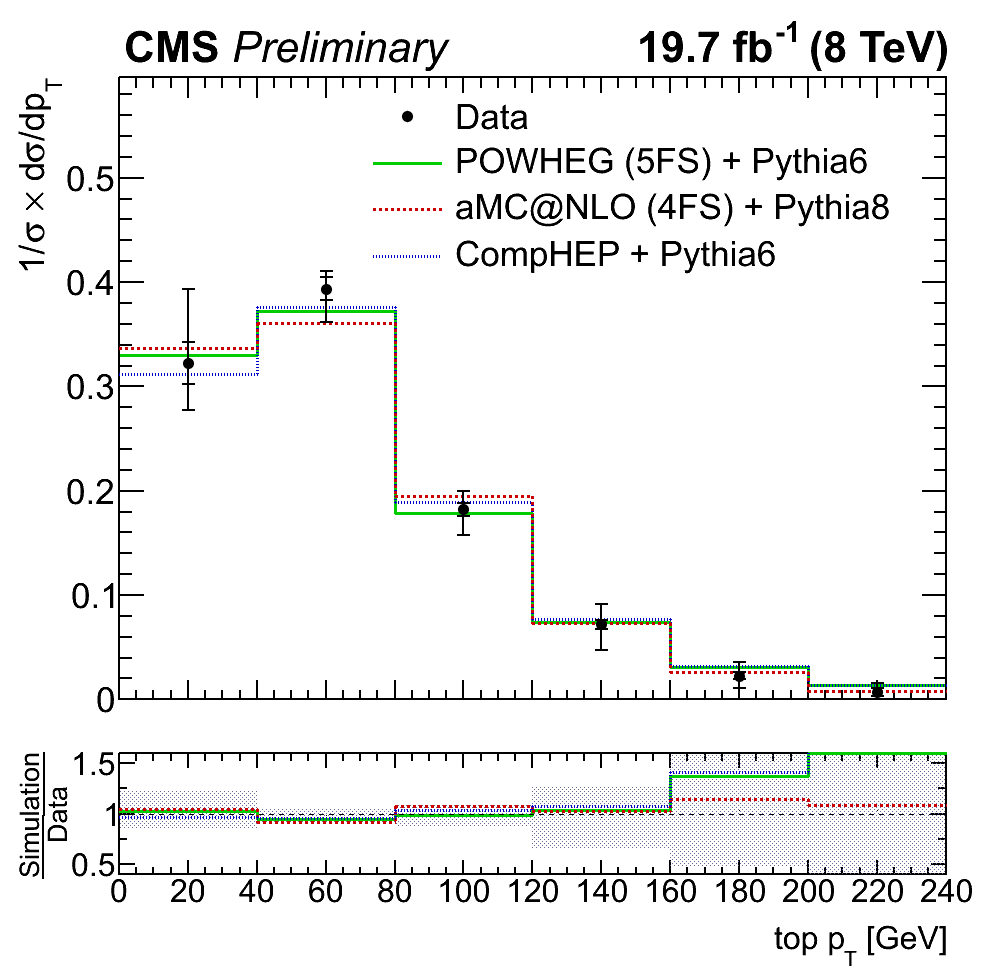} %
       \includegraphics[width=0.33\textwidth]{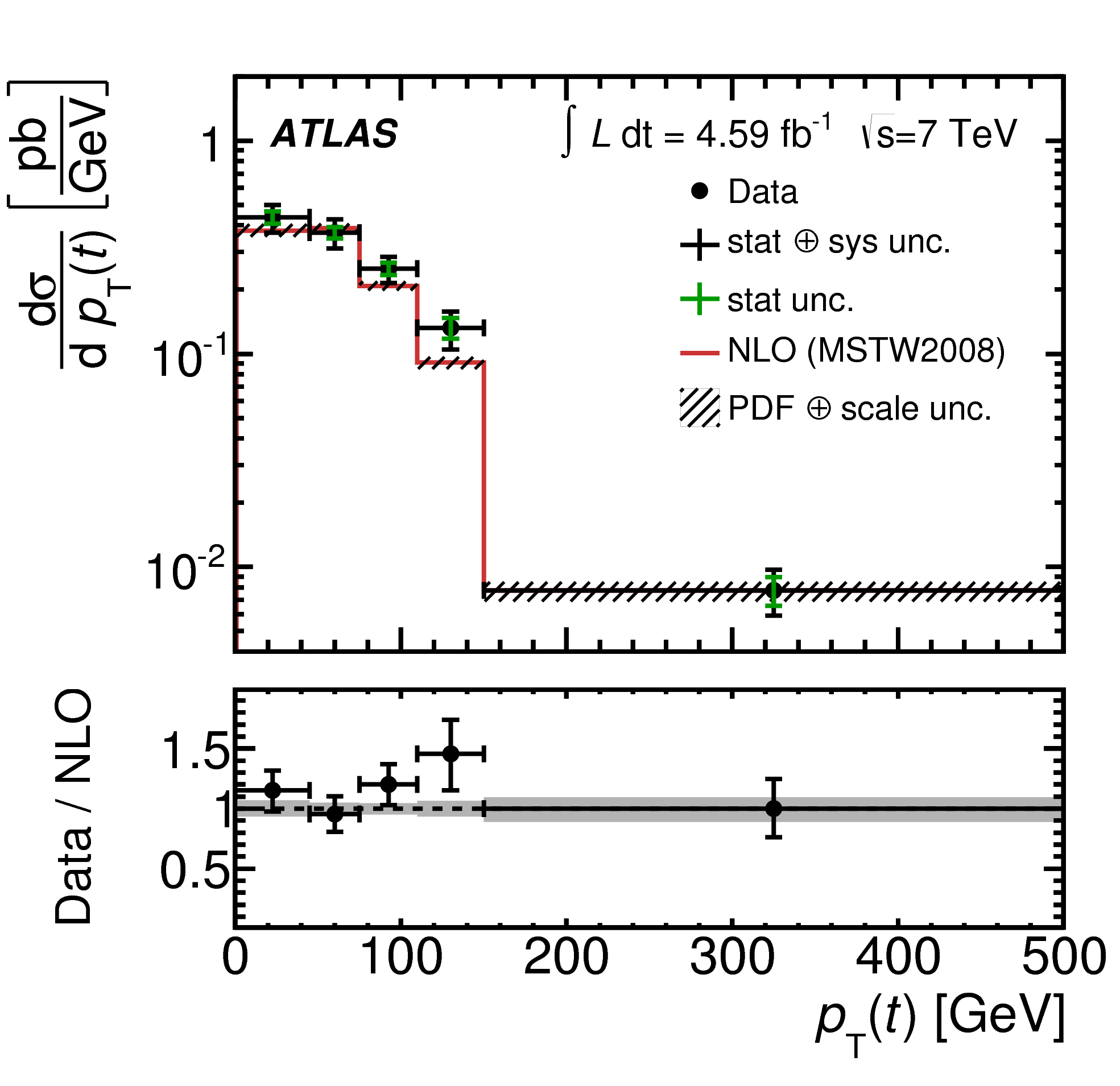}%
       \includegraphics[width=0.33\textwidth]{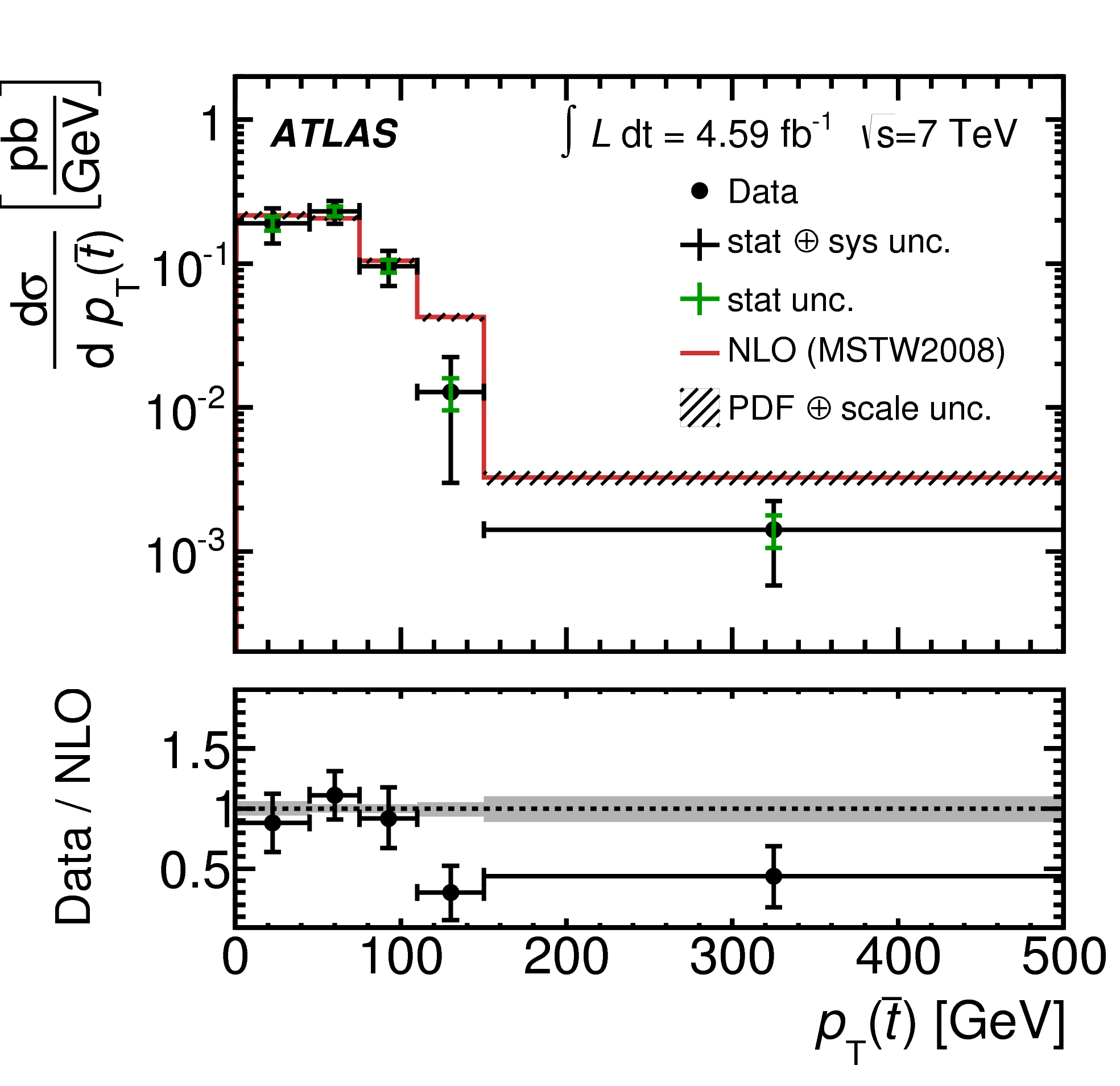}       
       \includegraphics[width=0.33\textwidth]{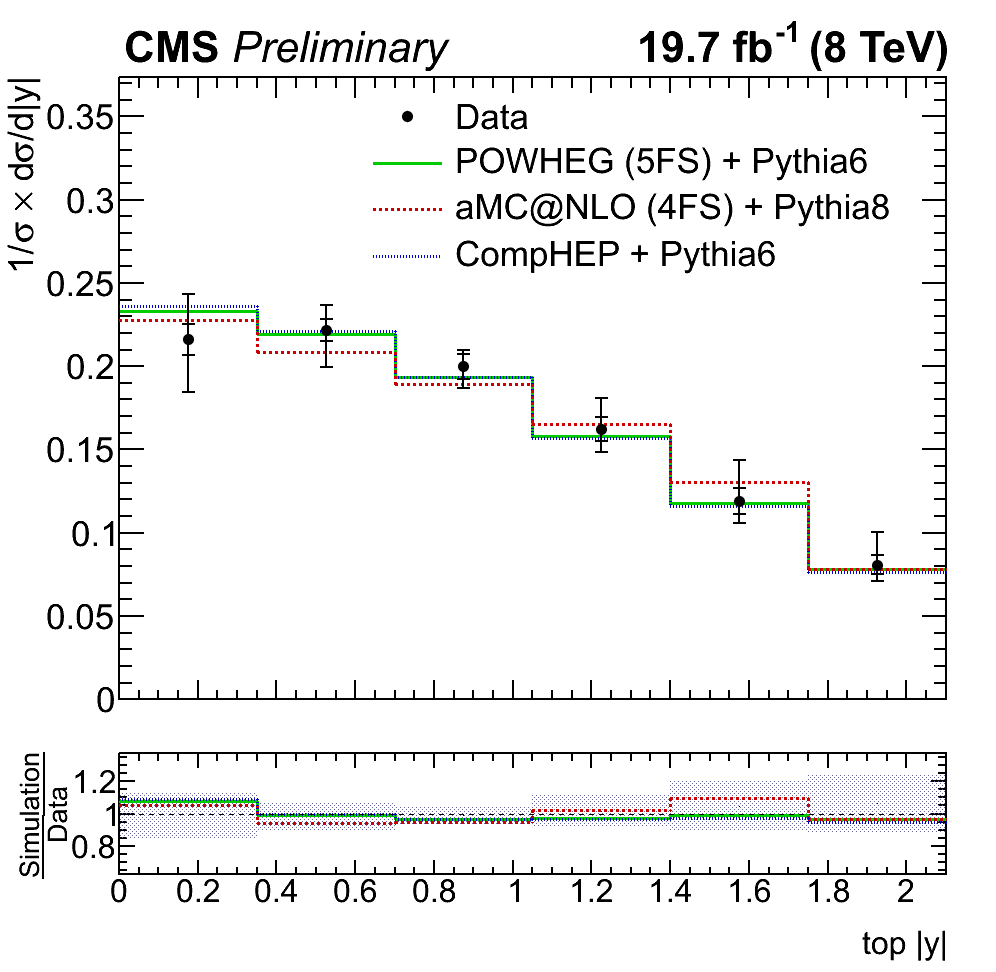}%
       \includegraphics[width=0.33\textwidth]{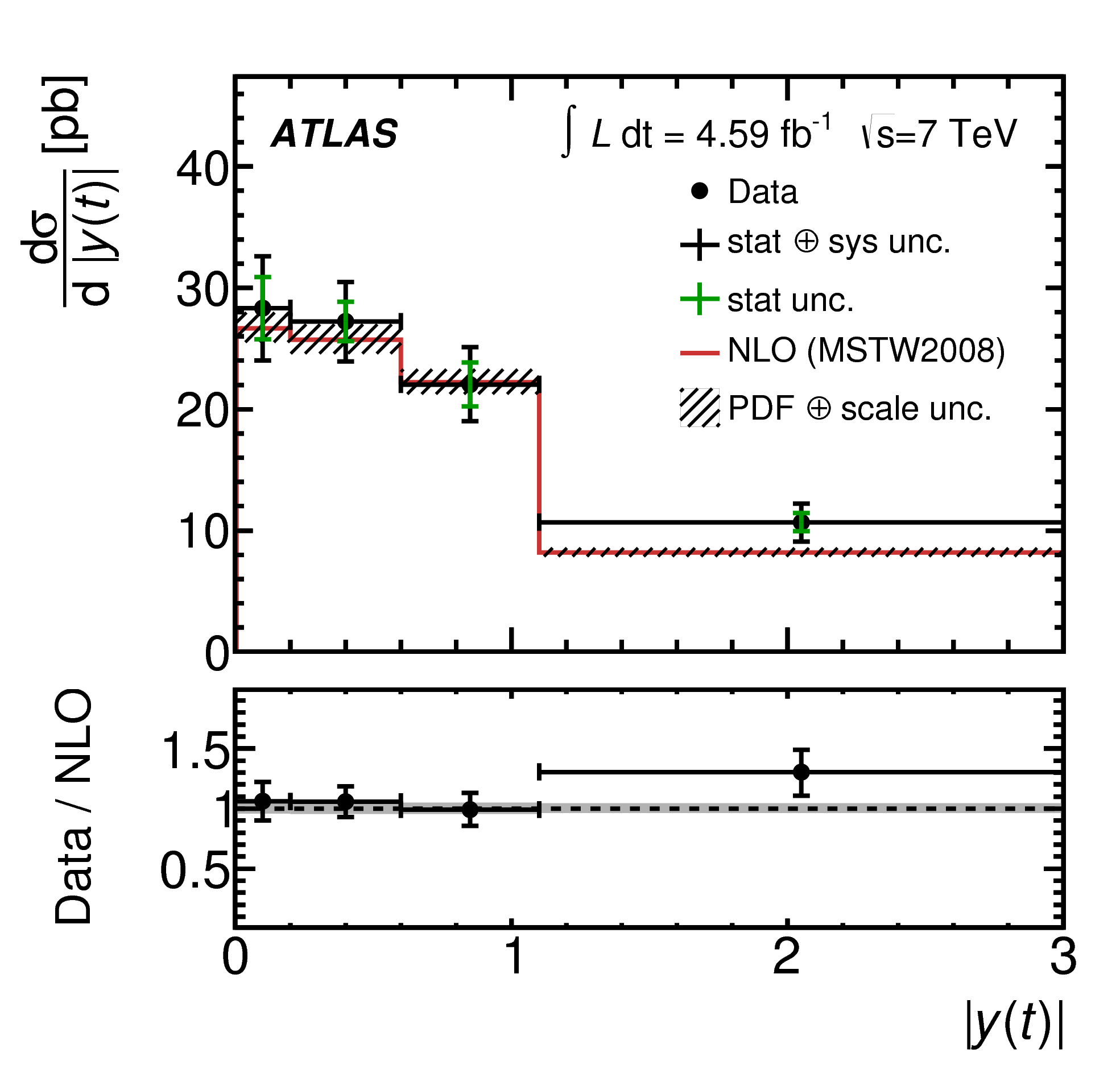}%
       \includegraphics[width=0.33\textwidth]{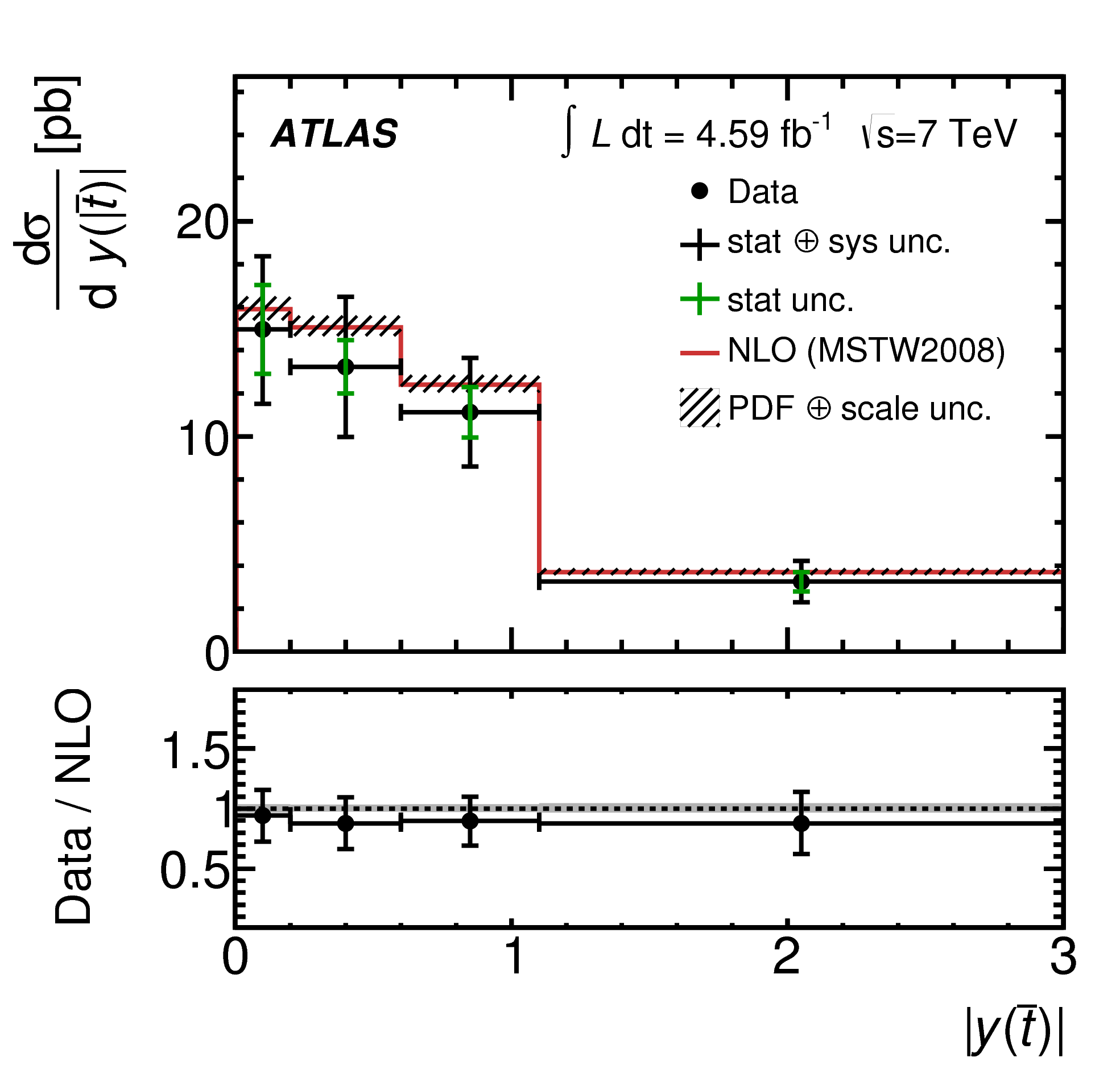}%
  \caption{Normalised differential cross section of the top and antitop quark production~\cite{CMSSingleTop} (left), and absolute $tq$ (middle) and $\bar{t}q$ (right) production cross sections~\cite{AtlasSingleTop} as a function of p$_T$ (top) and $|y|$ (bottom). The inner error bars indicate the statistical uncertainty while the outer error bars indicate the total uncertainty.} 
   \label{fig:single}
  \end{center}  
\end{figure}

Differential production cross section of single top quark t-channel as a function of the top quark p$_T$ and rapidity have been measured for the first time~\cite{AtlasSingleTop,CMSSingleTop}. In addition to the event selection and the kinematic reconstruction of the top quark, neural network techniques are used to further separate signal from background events. The measurements are as well corrected for selection efficiencies
and detector resolution effects with an unfolding technique. The normalised cross sections measured by CMS are shown in Fig.~\ref{fig:single}, left. The middle and right distributions correspond to the absolute $tq$ and $\bar{t}q$ production cross sections measured by ATLAS. The results are well described by the different predictions and they are in agreement with the expected $t/\bar{t}$ rates. The precision is generally dominated by the total systematic uncertainty. 

\section{Summary}

The differential $t\bar{t}$ production cross section in pp collisions has been measured as a function of several top quark related kinematic variables. The measurements are corrected to particle or parton level and are presented as normalised and absolute cross sections, in the full phase space or visible phase space, depending on the observable and the analysis. The results of $d\sigma/dp_T$ have been extended up to the 1 TeV regime. In general, the different predictions describe reasonably well the data. However, some discrepancies between data and certain predictions have been observed: the measured spectra are in some cases softer than predictions (ie. p$_T^t$) and the predicted cross sections tend to overestimate the data. Detailed comparisons between ATLAS and CMS verified the consistent definition of the top quark (parton level after radiation) and similar performance of the default generators. Measurements have been performed for the first time in single top (t-channel), both absolute and normalised. The results, dominated by the systematic uncertainties, are well described by the predictions. Overall, results are consistent among channels, measurements and experiments.

\end{document}